\documentclass[twocolumn,showpacs,preprintnumbers,amsmath,amssymb,pre,aps]{revtex4}
\usepackage{amsmath,amssymb}
\usepackage{epsfig}
\usepackage{graphicx}
\def\etal{\mbox{\it et al.}}

\newcommand{\be}{\begin{equation}}
\newcommand{\ee}{\end{equation}}

\begin{document}

\title{Phase-coexisting patterns, horizontal segregation and controlled convection in vertically vibrated binary granular mixtures}
\author{I. H. Ansari$^1$, N. Rivas$^{2,3}$ and M. Alam$^{1*}$}
\affiliation{
$^1$Engineering Mechanics Unit, Jawaharlal Nehru Centre for Advanced Scientific Research, Jakkur PO, Bangalore 560064, India\\
$^2$Multi-Scale Mechanics, MESA+, University of Twente, Netherlands\\
$^3$Forschungszentrum J\"ulich GmbH, Helmholtz-Institut Erlangen-N\"urnberg  f\"ur Erneuerbare Energien (IEK-11), F\"urther Strasse 248,
90429 Nuremberg, Germany\\
$^*$Email for correspondence: meheboob@jncasr.ac.in\\
}

\date{\today}

\begin{abstract}
We report new patterns, consisting of  coexistence of sub-harmonic/harmonic and asynchronous states  [for example, a granular gas co-existing with (i)  bouncing bed, (ii)  undulatory subharmonic waves and (iii)  Leidenfrost-like state], in experiments  on vertically vibrated binary  granular mixtures in a Heleshaw-type cell. Most experiments have been carried out with equimolar binary mixtures of glass and steel balls of same diameter by varying the total layer-height ($F$) for a range of shaking acceleration ($\Gamma$). All patterns as well as the related phase-diagram in the ($\Gamma, F$)-plane have been reproduced via molecular dynamics simulations of the same system. The segregation of heavier and lighter particles along the horizontal direction is shown to be the progenitor of such  phase-coexisting patterns as confirmed in both experiment and simulation. At strong shaking we uncover a {\it partial} convection state in which a pair of convection rolls is  found to coexist with a Leidenfrost-like  state. The crucial role of the relative number density of two species   on controlling  the buoyancy-driven granular convection is demonstrated. A possible model for spontaneous horizontal segregation is suggested based on anisotropic diffusion.
\end{abstract}


\maketitle

\section{Introduction}

Real particles are always polydisperse, the simplest case being a binary mixture of particles of different density and/or size.
A binary granular mixture driven by external vibrations is known to  lead to spontaneous
segregation or demixing~\citep{Brown1939} of two species, which can be a nuisance/blessing for industries dealing with granular materials.
The  segregation  in binary mixtures has been extensively studied under vertical/horizontal  vibration~\citep{Kudrolli2004} 
as well as in different shearing experiments~\citep{OK2000},
although a unified theory for Brazil-nut segregation~\citep{Rosato1987} and  its reverse~\citep{SM1998,HQL2001,Breu2003} is still missing.
Understanding segregation-induced patterns can help to  control and manipulate particulate flows, for example, in chemical and pharmaceutical industries.
Moving  away from applications, patterns are important in dynamical systems theory and non-equilibrium statistical mechanics, 
since they provide an avenue to probe the validity of any self-consistent theory. 
Understanding segregation-driven patterns and finding ways to control them is a key challenge in granular physics research.

A vertically shaken box of particles displays a rich variety of patterns: Faraday-type sub-harmonic waves and heaping~\citep{Faraday,DFL1989,CDR1992},
oscillon~\citep{UMS1996},  convection~\citep{GHS1992,WHP2001,Eshuis2007,Eshuis2010},
Leidenfrost-like density inversion~\citep{MPB2003,Eshuis2005} and segregation~\citep{PH1995,BKS2002},
with each pattern being characterized by its distinct spatial and temporal symmetries.  
While patterns having different spatial symmetry are routinely found  to coexist (e.g.~the gas, liquid 
and solid  coexists in driven granular matter~\citep{OU1998,AT2006}),
the coexistence of patterns with different temporal symmetry (e.g.~a sub-harmonic-wave  coexisting with a harmonic or an asynchronous state) is  rare.
An example of latter-type patterns is the well-known  ``oscillon''~\citep{UMS1996} which was discovered in Faraday-type experiments
on  granular materials under vertical shaking -- this represents a period-2 state that coexists with a period-1 bouncing-bed.
Such patterns having different spatial and temporal symmetries are heretofore dubbed `phase-coexising' patterns.
For  monodisperse  granular particles under harmonic shaking, we refer to recent works in Ref.~\cite{Eshuis2007,SAMLA2014,AA2016} for a broad overview of all observed patterns
as well as contributions of different groups; the present work is focussed on uncovering patterns in binary granular mixtures in a Heleshaw-type container
under vertical shaking.

In this  paper, a variety of new patterns, having {\it different spatial and temporal symmetries} (one example being a period-2 wave coexisting with
a disordered/asynchronous granular gas), is uncovered  in vertically shaken binary granular mixtures that challenges current theoretical understanding of granular flows.
It is shown that the  ``buoyancy-induced'' granular convection can be controlled by the addition of
a small amount of second species, leading to a ``partial-convection'' state in which the convection-rolls span
only a part of the Heleshaw-cell, coexisting with a  Leidenfrost-like state in the remaining part of the cell.
The genesis of these novel patterns is shown to be  tied to the segregation of two species along the horizontal direction
which is also confirmed  in particle-dynamics simulations.  
A possible driving mechanism for horizontal segregation under purely vertical shaking
is discussed in terms of an anisotropic diffusion tensor.

\section{Experimental method and simulation technique}
 
The experimental setup consists of a quasi-two-dimensional rectangular Plexiglas container which is mounted on an electromagnetic shaker
(Ling Dynamics System V456, Br\"uel \& Kjaer).
The scaled length ($L/d$), depth ($D/d$) and height ($H/d$) of this container are $100$, $5.5$ and  $80$, respectively;
a similar setup was used by Eshuis {\it et al.}~\cite{Eshuis2007} and subsequently by two of the present authors~\citep{AA2013,AA2013a,AA2016} to probe
the pattern-formation scenario in a vertically driven mono-disperse granular matter.
In this paper we consider a  binary mixture of spherical glass and steel balls of  the same diameter $d\approx 1.0$ mm, 
having a density ratio of $\rho_{steel}/\rho_{glass} \approx 3.06$.
The container is  filled up-to a specified layer-depth at rest,
\begin{equation}
    F=\frac{h_T}{d} =\frac{h_g}{d} + \frac{h_s}{d} \equiv F_g + F_s \in (2.5,\ldots 10),
 \label{eqn:eqn2}
\end{equation} 
where $h_g$ and $h_s$ are the initial depths of glass and steel balls.
An equimolar ($50:50$) mixture of same-size particles, for which the initial layer-heights of glass and steel balls are equal ($h_g=h_s$), was used in most experiments;
the effect of varying the relative number fraction of steel balls [$F_s/F=h_s/h_T\in (0,0.5)$]  was also assessed for few cases.

The particle-filled container is vibrated vertically, with a harmonic wave $y=A\sin{(2\pi f t)}$ of amplitude $A$ and frequency $f$
by  the electromagnetic shaker which operates in a closed-loop, controlled by a 
controller (COMET, B\&K) and an   amplifier (PA 1000L, B\&K) through a  software interface. 
To generate a feedback signal of required amplitude and frequency of the harmonic excitation, 
a piezoelectric accelerometer (DeltaTron$^{\copyright}$) is attached on the head-expander over which the Plexiglas container is mounted.

\subsection{Initial configuration and experimental protocol}

The experiments were performed for a range of shaking acceleration/intensity 
\begin{equation}
    \Gamma = \frac{4\pi^2Af^2}{g} \in(0, 50),
\label{eqn:eqn1}
\end{equation}
in both {\it up-sweeping} (increasing frequency $f$  from the rest state) and {\it down-sweeping} 
(decreasing frequency from an excited state at the same ramping rate) modes.
For a specified shaking amplitude $A/d$ ($= 3, 6$), the shaking acceleration 
$\Gamma$, (\ref{eqn:eqn1}), was increased/decreased by increasing/decreasing frequency ($f$)
at a linear rate; all results presented corresponds to a linear frequency-ramping of $0.01\, Hz/s$;
the results are found to be almost identical for a ramping rate of $0.1$ Hz/sec.
To assess the stability of observed patterns, many experiments were repeated at fixed $\Gamma$ for at least 30 minutes
 [$>O(10^4\tau)$, where $\tau=1/f$ is the time-period of shaking].

The up-sweeping experiments were done with an initial configuration of randomly mixed mixture of steel
and glass balls, and the final state of each up-sweeping experiment
is used as the initial state for its {\it down-sweeping} evolution at the same rate of frequency ramping.
Some experiments were also repeated with {\it segregated} initial states:
(i) steel balls on top of glass balls and (ii) glass balls on top of steel balls;
the  reported patterns are found to be qualitatively similar irrespective of the initial state.

\subsection{Imaging and velocity measurement}

The temporal evolution of the collective motion of particles has been recorded 
using a high-speed camera (IDT MotionPro Y4S3) at $1000$ frames/second;
this camera can operate at $5100$ fps  at the full resolution of $1016\times 1016$ pixels.
The images were analyzed to yield  information on the onset value of $\Gamma$ for different types of patterns 
and their characteristics (sub-harmonic or synchronous, segregation,  and coexistence of different phases).
Moreover, a commercial PIV (Particle Image Velocimetry) software (``Dynamic Studio Software''  of Dantec Dynamics A/S, Denmark)
was used to extract coarse-grained/hydrodynamic velocity field by processing the images using an adaptive-correlation technique
in which the size of the interrogation window is varied adaptively from $64\times 64$ to $16\times 16$ pixels, with  $50\%$  overlap.
The related experimental details  of image-analysis and the PIV methodology are documented in Ref.~\cite{AA2016}.

\subsection{Event-driven simulation}

Event-driven simulations for a similar system of vibrated binary mixture were  carried out
to cross-check the robustness of the present  experimental findings.  A  binary mixture of equal size
hard inelastic spherical particles of a specified density-ratio ($\rho_{steel}/\rho_{glass}=3.06$) is used as a model granular mixture. 
The particles are considered as rough hard spheres with translational
and rotational degrees of freedom, and the whole box is vibrated by a bi-parabolic sine-interpolation in the same region of the phase-space of experiments;
the event-driven algorithm and simulation details can be found in Ref.~\cite{Rivas2011}.
The collisions between particles are modelled by normal and tangential restitution coefficients:
$e_n=e_t=0.95$ for glass-glass,  $e_n=e_t=0.85$ for steel-steel and  $e_n=e_t=0.9$ for steel-glass collisions;
the static and dynamic friction coefficients are $\mu_s=\mu_t=0.1$ for both materials.
The collisions of particles with the bottom-plate as well as with front, back and two side
walls are modelled same as particle-particle collisions, with $e_n=e_t=0.95$ and $\mu_t=\mu_d=0.1$. 
We have checked that replacing the front- and back-walls with periodic boundary conditions do not affect the observed patterns.

\begin{figure}
\begin{center}
(a) \includegraphics[width=3.3in]{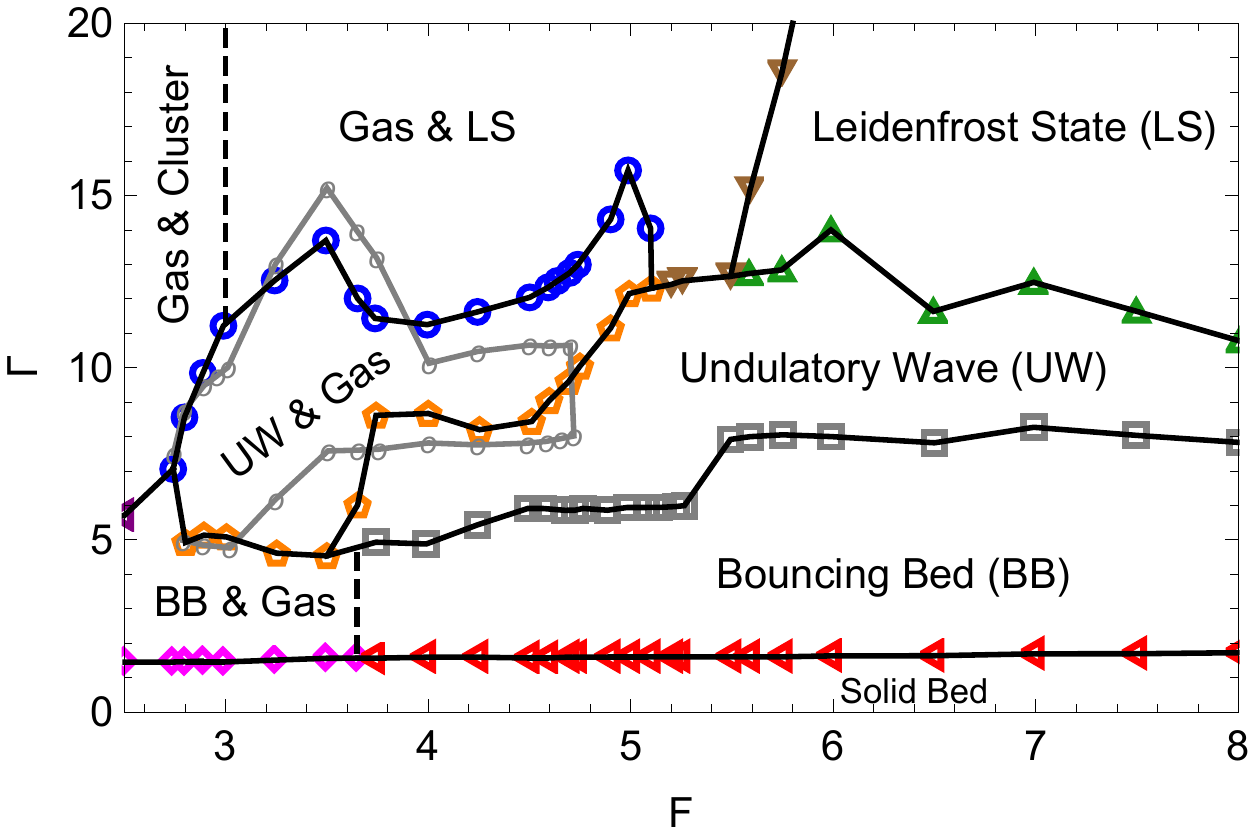}\\
(b) \includegraphics[width=3.3in]{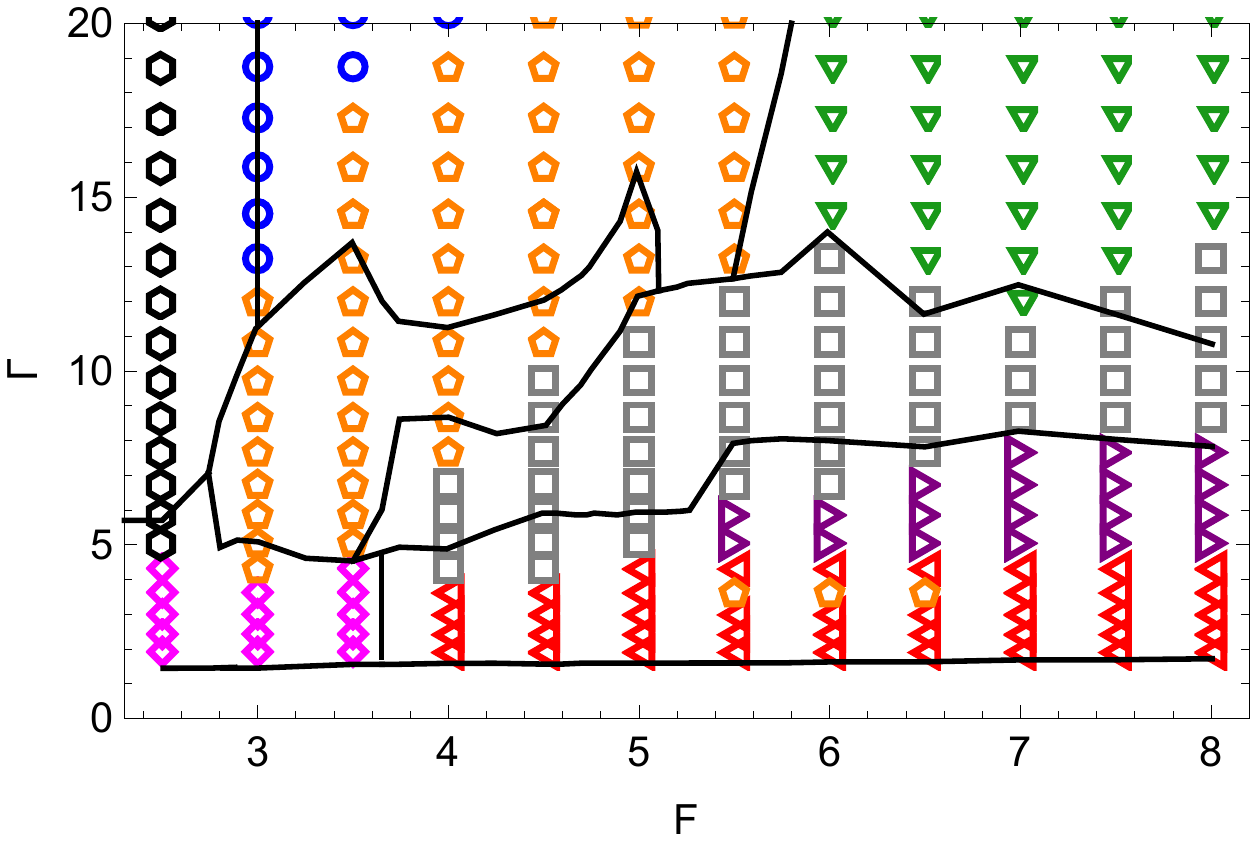}
\caption{
(a) Phase diagram of patterns in ($\mathbf\Gamma, F$)-plane for up-sweeping experiments for $F_g=F_s$ (equimolar mixture) with a shaking amplitude of $A/d=3$.
(See Movies 1 and 2 for a visual inspection of different patterns.)
The symbols represent the locations of transitions between two states (e.g.,~`BB+Gas' and `UW+Gas') for a specified layer depth $F$; 
the vertical dashed line represents an approximate phase-boundary between two consecutive $F$ that represent two different states (~`Gas+Cluster' and `Gas+LS');
the data for down-sweeping experiments (open grey symbols) for the region of `UW+Gas' patterns are also superimposed; see text for details.
(b) Phase-diagram from MD-simulations; the broken black-lines denote the approximate phase-boundaries obtained from up-sweeping experiments (same as in panel $a$);
 refer to the text for patterns related to different symbols.
}
\label{fig:fig1}
\end{center}
\end{figure}

\section{Patterns, segregation and convection control}

\subsection{Phase diagram of patterns: experiment vs simulation} 

We begin by describing the complete phase diagram of all patterns  in the ($\Gamma, F$)-plane for up-sweeping experiments, 
shown in Fig.~\ref{fig:fig1}(a); we also refer to   Supplementary Movies 1 and 2 for a visual inspection of different patterned states.
The symbols in Fig.~\ref{fig:fig1}(a) represent the onset of a new state/pattern when the shaking intensity ($\Gamma$) is increased from zero for a specified layer depth $F$;
the broken-lines joining different symbols represent `approximate' phase-boundaries for the transition between two patterned-states in the ($\Gamma, F$)-plane.
The stable nature of different types of patterns, marked in Fig.~\ref{fig:fig1}(a), has been verified by repeating experiments
at selected values of $\Gamma$ and $F$ over a long time $t/\tau>10^3$.

For $\Gamma\sim 1$, the kinetic energy of particles is not able to overcome their dead-weight 
and hence the granular bed moves synchronously with the container without getting detached  from the base plate -- this is called  {\it solid bed}, 
that occurs below the  horizontal line around $\Gamma\sim 1$ in  Fig.~\ref{fig:fig1}(a).
With increasing $\Gamma>1$, the bed detaches from the base  and starts bouncing 
like a single body, akin to a particle bouncing off a plate with zero restitution coefficient -- this is the regime of {\it bouncing bed} (BB) 
whose motion is synchronized with the external shaking frequency  and hence called a {\it synchronous} wave (see  Movie-1).
For the present system of a binary mixture, the solid bed gives birth to a  BB state for $F \geq  3.75$, 
and a mixed `{\it BB \& Gas}' state for smaller filling heights $F<3.75$ -- these two regions are marked in Fig.~\ref{fig:fig1}(a) above the solid-bed region.

\begin{figure}
\begin{center}
\includegraphics[width=3.4in]{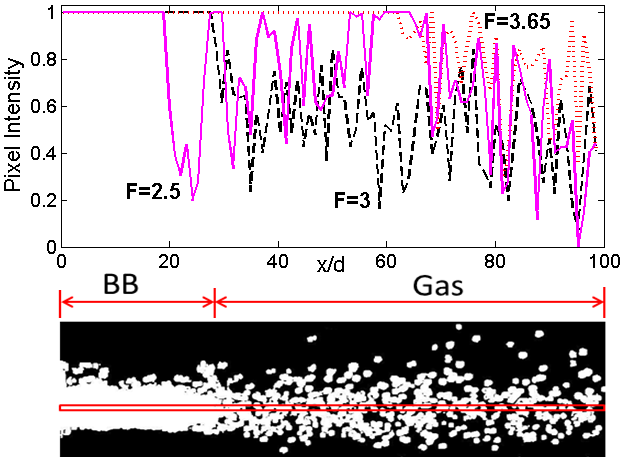}
\caption{
Pixel-intensity profiles (upper panel) for patterns at $\Gamma=3.5$ for different $F$ (refer to Fig.~\ref{fig:fig1}); the lower panel shows the binary image of the pattern for $F=2.5$,
and thin red-box represents the region over which the pixel-intensity is vertically-averaged to yield  the profiles shown in the upper panel.
}
\label{fig:fig2}
\end{center}
\end{figure}

An illustration of the   `{\it BB \& Gas}'  state is displayed in the lower panel of Fig.~\ref{fig:fig2} at a shaking intensity of $\Gamma=3.5$ for a layer-height of $F=2.5$.
This represents a binary pixel-image of the snapshot of the instantaneous particle configuration -- it is seen that
a relatively low-density state (populated largely by heavier steel balls) on the right of the container coexists with a bouncing bed state on its left.
The upper panel of Fig.~\ref{fig:fig2} displays  the average pixel-intensity  profiles  for  $F= 2.5$, $3$ and $3.65$--
these profiles have been calculated by averaging the pixel intensities over a horizontal strip (see the red-colored box in the lower panel) of height of approximately two particle diameters.
It is clear that with increasing $F$ the length of the dilute gaseous region decreases,
with a transition from `{\it BB \& Gas}' to a complete BB-state occuring at $F>3.75$ (marked by the vertical dashed line in Fig.~\ref{fig:fig1}a for $\Gamma<5$).

In  Fig.~\ref{fig:fig1}(a) the region enclosed by blue circles and orange pentagons represents another  ``mixed'' pattern (``UW \& Gas'') 
consisting of a period-2 undulatory wave (UW) and a  gas-like state --  in this region the data for down-sweeping experiments, marked by grey circles, are also superimposed.
These experiments were repeated at least three times, and  we found hysteretic-like behaviour for the onset of `UW+Gas' pattern only at larger layer depths 
i.e.~ $\Gamma_{onset}^{up} > \Gamma_{onset}^{down}$ for $F>4.0$. 
(For other patterns marked in Fig.~\ref{fig:fig1}(a), the onset value of $\Gamma$ at any $F$ shows little variation for upsweeping and down-sweeping experiments.)
From Supplementary Movie-1, it can be verified   that the mixed state of ``UW \& Gas'' 
maintains its shape and position for a very long time, and hence this as well as other patterns in Fig.~\ref{fig:fig1}(a)  are stable.
The  characteristic features and novelty of ``UW \& Gas''  pattern and its possible origin are discussed later in Sec~III.B.

To check the robustness of new patterned-states in Fig.~\ref{fig:fig1}(a), we have  constructed the phase-diagram of patterns based on MD-simulations as shown in Fig.~\ref{fig:fig1}(b);
the symbols represent simulation data that were obtained by running simulations at specified values of layer-depth $F$
and ramping-up shaking frequency (and hence increasing $\Gamma$) as done in experiments.
It is clear that all patterns uncovered in experiments are also found in simulations, although there are variabilities in terms of
the range of ($\Gamma, F$) over which each pattern can persist  in simulations. For example, the value of $\Gamma$ for the onset of the
mixed-state ``UW \& Gas'' (denoted by orange-colored pentagons in panel $b$) agrees closely with experiments,
but these patterns persist at much larger values of $\Gamma$ in simulations than in experiments.
Consequently, another mixed-pattern of ``LS+Gas'' (the  Leidenfrost-like state (LS)~\citep{Eshuis2005} coexisting with a gas-like state; 
denoted by blue circles in Fig.~\ref{fig:fig1}(b)) are found at much larger values of $\Gamma$ in simulations.
The remaining patterns in Fig.~\ref{fig:fig1}(b),  such as  ``Gas+Cluster'' (black hexagons), ``BB+Gas'' (pink diamonds),
``BB'' (denoted by left triangles),  ``UW'' (grey squares), and ``LS'' (inverted green triangles) are found to encompass 
almost the same regions of $(\Gamma, F)$-space in simulations and experiments.
A detailed description of  the ``LS \& Gas'' pattern  is deferred to Sec.~III.C, and the
onset of convection (that occurs at much higher values of $\Gamma$, not shown in Fig.~\ref{fig:fig1}, but see supplementary movies as well as Fig.~\ref{fig:fig8}) 
and its control are discussed  in Sec.~III.D.
On the whole, the comparison between experiment and simulation  in Fig.~\ref{fig:fig1}(b) confirms that  all patterns uncovered in experiments are robust.

\begin{figure}
\centering
(a) \includegraphics[width=3.1in]{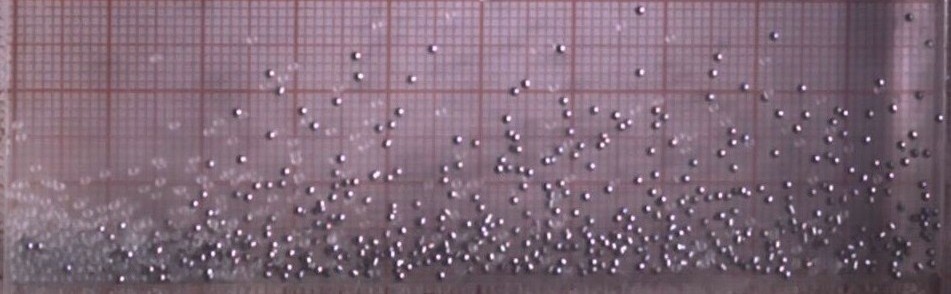} \\
(b) \includegraphics[width=3.1in]{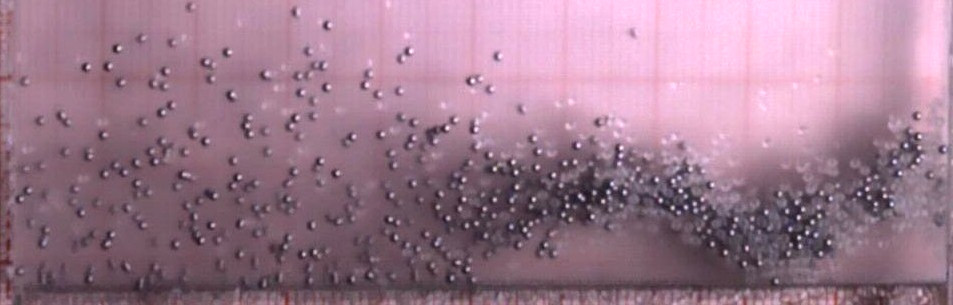}\\
(c) \includegraphics[width=3.1in]{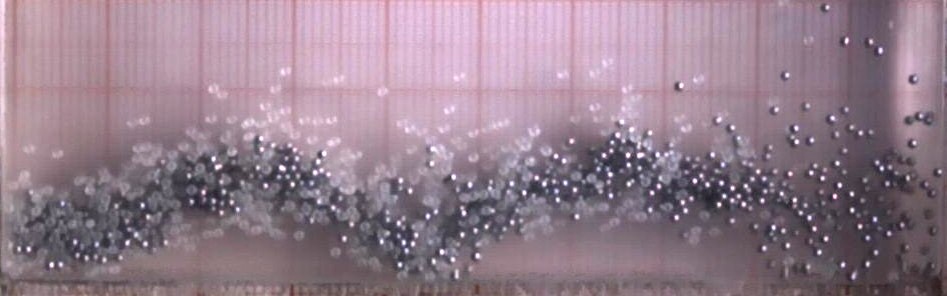}\\
(d) \includegraphics[width=3.1in]{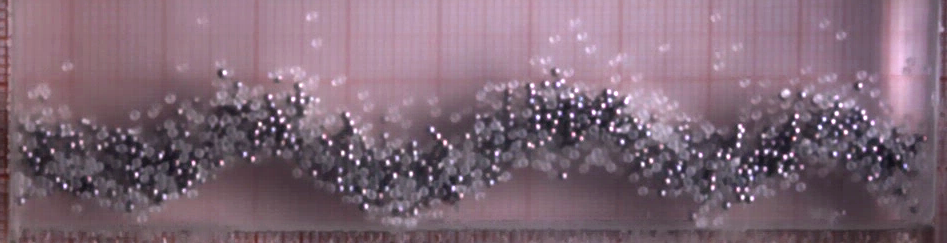}
\caption{
Emergence of different patterns with increasing filling depth ($F$) at $\Gamma\approx  9$.
(a) `{Gas \& Cluster}' state at $F =2.5$; (b,c) ``Undulatory Wave (UW)  \& Gas'' at ({b}) $F =3$  and ({c}) $F =4$ ; (d) complete UW  at $F =5$.  
The dark and light grey particles correspond to steel and glass balls, respectively.
}
\label{fig:fig3}
\end{figure}

\subsection{ Coexistence of synchronous and asynchronous states: ``UW \& Gas'' and ``Gas \& Cluster''}

Figure~\ref{fig:fig3} displays four images with increasing total filling depth ($F=F_g+F_s$) at  $\Gamma\approx 9$, refer to Fig.~\ref{fig:fig1}.
A cluster of glass-rich balls is seen on the left of the container in Fig.~\ref{fig:fig3}(a) that coexists with a gas-like state of heavier steel balls on the right
of the container at $F_g=F_s=1.25$ -- this is dubbed ``Gas \& Cluster'' pattern (which is marked in the upper-left side in the ($\Gamma, F$)-plane in Fig.~\ref{fig:fig1}); 
the cluster was also observed on the right side of the container.
In either case, once formed, the cluster retains its position for a long time exceeding O($10^4\tau$) and therefore the ``Gas \& Cluster'' is a stable pattern.
At the same $F=2.5$ but with increasing $\Gamma$, the cluster size decreases, it undergoes a random oscillation  and eventually vapourizes
to give birth to a granular gas at large enough $\Gamma>40$ (not shown).

When the total layer depth is increased to $F=3$,  we find  a coexisting  pattern of (i) a granular gas on the left of the container 
and (ii) an undulatory wave (UW)  on its right as shown in Fig.~\ref{fig:fig3}(b) -- this is dubbed ``UW \& Gas'' pattern.
For even deeper beds ($F=5$),  a complete UW  is found to span the whole length of the container, see Fig.~\ref{fig:fig3}(d).
These UWs are {\it sub-harmonic} standing waves whose time-period of oscillation is twice the time-period ($\tau=1/f$) of shaking  
(i.e.~the peaks and valleys of UWs repeat after $2\tau$, see Movie 1) and hence  called `$f/2$' or `period-2' waves.
A `partial' UW state, such as  Figs.~\ref{fig:fig3}($b,c$),
maintains its shape and position for a very long time ($t> 5\times 10^4\tau$) -- hence these patterns are indeed stable and long-lived.

While the `{\it Gas \& Cluster}' state (Fig.~\ref{fig:fig3}a) represents the coexistence  
of two phases having different spatial-order (gas and liquid) and different temporal symmetry,
the `{\it UW \& Gas}' state (Figs.~\ref{fig:fig3}b,c) represents a rare coexistence  of a period-2 sub-harmonic wave and an asynchronous/disordered gas-like state. 
Such coexisting states of subharmonic/harmonic and asynchronous/disordered states
having different spatial and temporal symmetries are heretofore dubbed {\it phase-coexisting} patterns.

In the context of vibrated ``monodisperse'' granular materials, the experimental work of Gotzendorfer~{\it et al.}~\cite{GKRS2006}
first reported the coexistence of subharmonic/harmonic and asynchronous (gas-like)  states. However these experiments were carried out
under ``combined'' vertical and horizontal shakings; the observed coexistence of gas and subharmonic bouncing state 
disappeared when the horizontal component of excitation is switched off.
Another related pattern is  {\it oscillon} \citep{UMS1996}  which represents a period-2 liquid-like state that 
coexists with a period-1 solid-like bouncing-bed state;  since two parts of an oscillon are  temporally synchronous  (with different time-periods) as well as spatially ordered,
 the oscillons must also be  categorized as different from the presently uncovered phase-coexisting patterns.
  In contrast to above two works, the present experiments deal  with binary granular mixtures that show coexisting patterns
  with spontaneous horizontal segregation (Fig.~\ref{fig:fig3})  subject to  harmonic excitation along the vertical direction.
  Under purely  vertical vibration, we are unaware of any work where the coexistence of  ``synchronous'' and ``asynchronous'' states  were reported
 in either monodisperse/binary mixtures.
The  onset of coexisting synchronous and asynchronous  patterns seems to be tied to the time-evolution of the segregation of two species 
as we demonstrate below via particle-dynamics simulations of our experimental setup.

\begin{figure}
\begin{center}
(a)\includegraphics[width=3.2in]{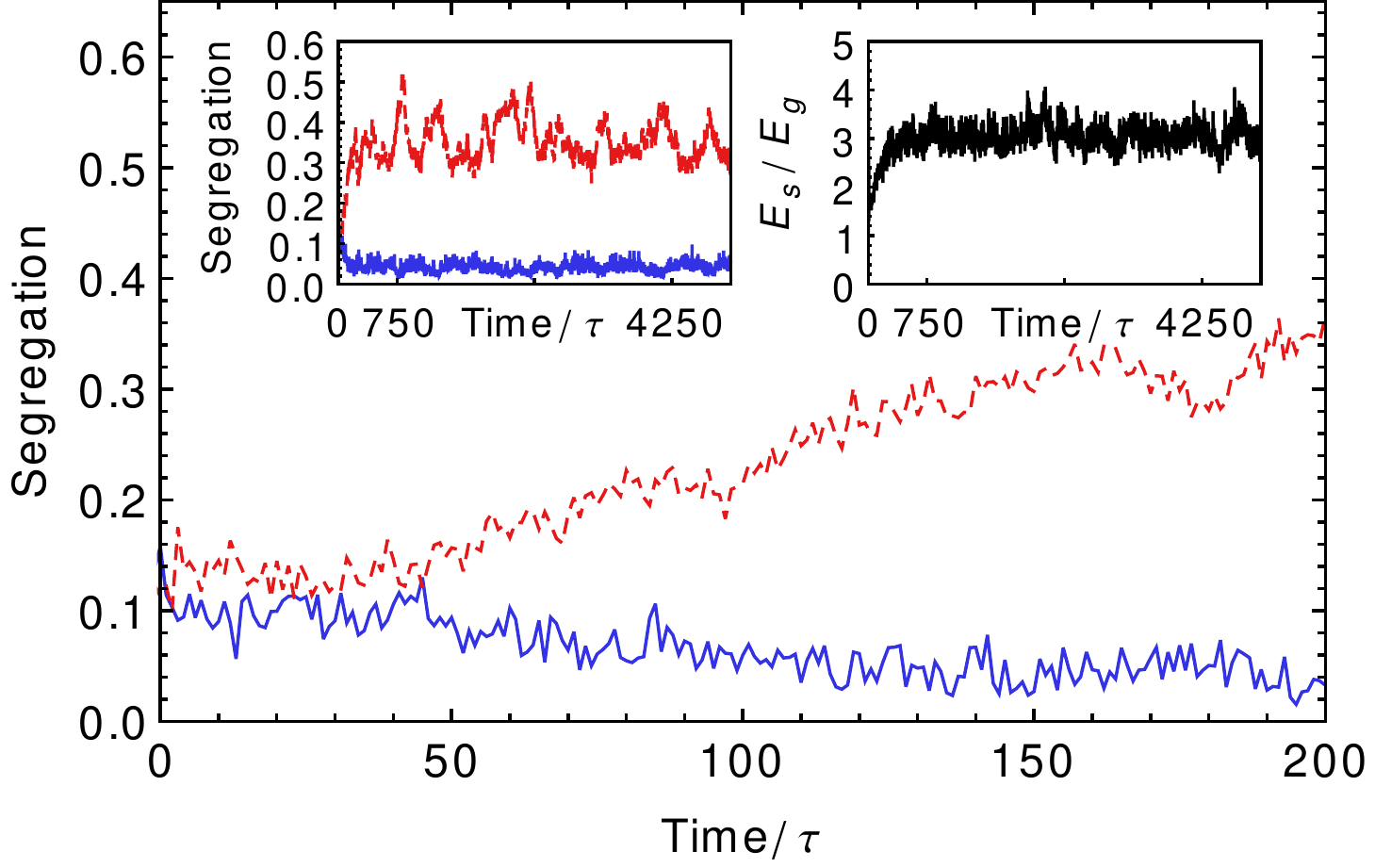}\\
(b) \includegraphics[width=3.1in]{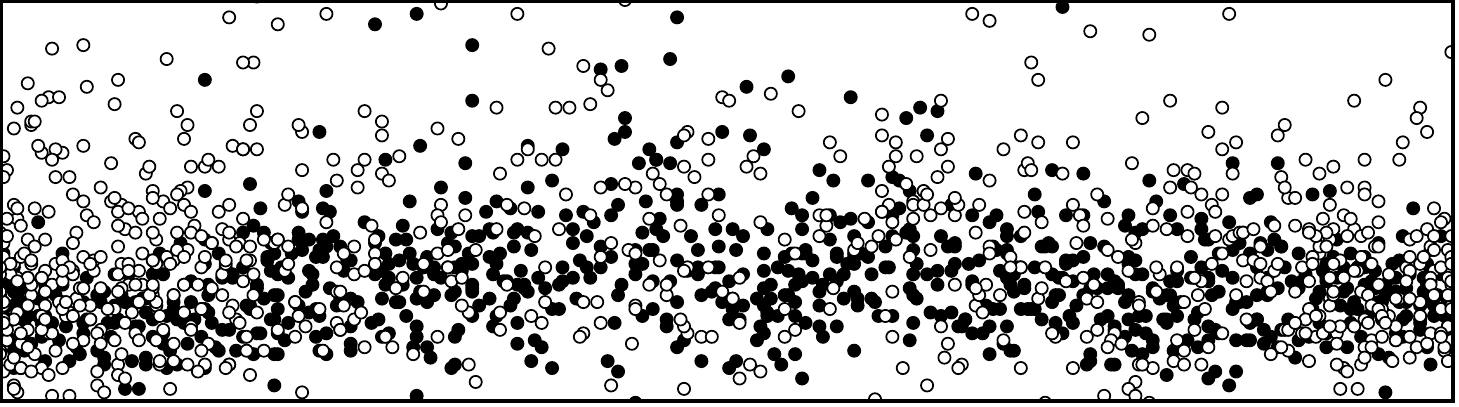}\\
(c) \includegraphics[width=3.1in]{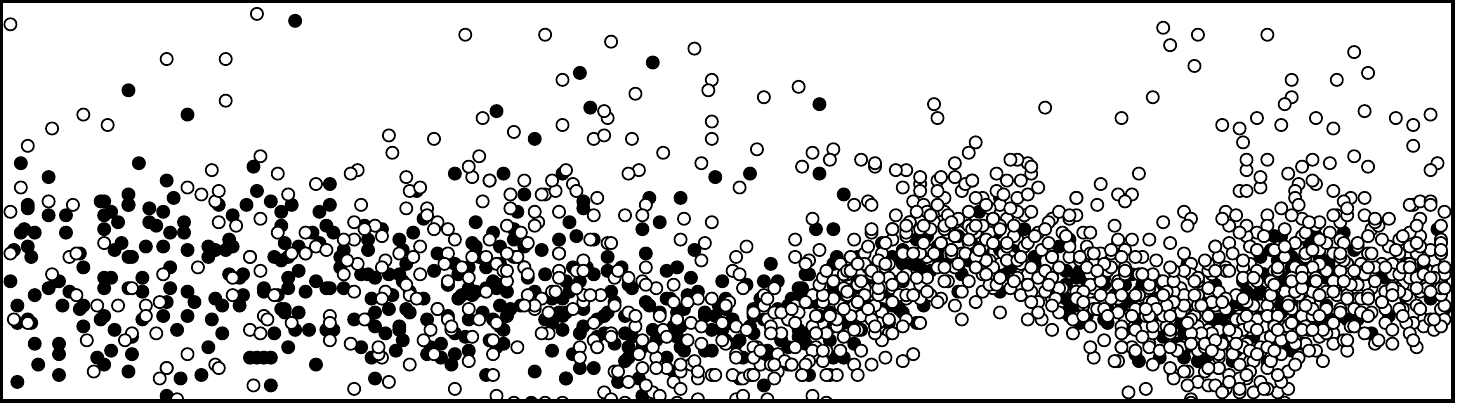}\\
(d) \includegraphics[width=3.1in]{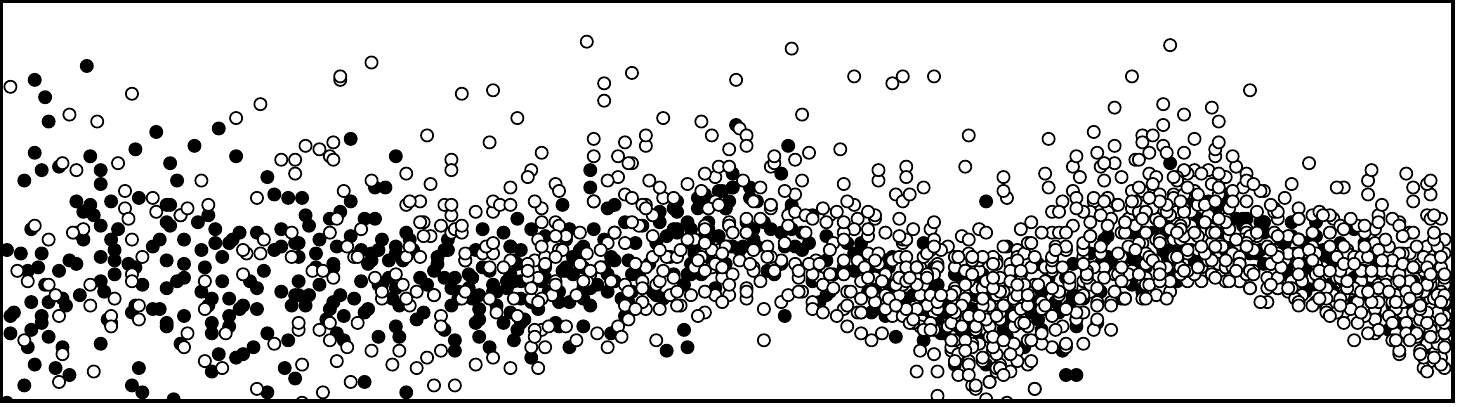}\\
(e) \includegraphics[width=3.1in]{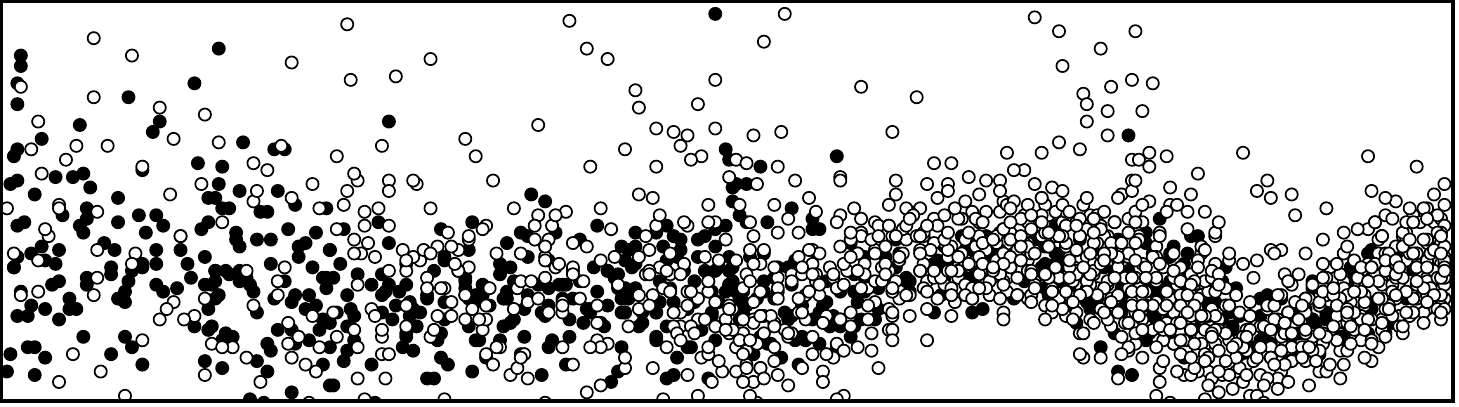}
\caption{
Simulation data for (a) the temporal evolution of  horizontal (red line)  and vertical (blue line) segregation  index:
short- and long-time evolutions are in the main panel and the left inset, respectively.
Snapshots of particles at (b) $t=100\tau$, ($c$) $2500\tau$, ($d$) $2500\tau + \tau/2$ and ($e$) $2501\tau$; filled and open circles refer to steel and glass balls, respectively.
Right inset in panel ${a}$ shows the evolution of energy ratio, $E_s/E_g = \langle m_s\vec{C}_s^2\rangle/\langle m_g\vec{C}_g^2\rangle$. 
Parameters are as in experiments of Fig.~\ref{fig:fig3}(b).
}
\label{fig:fig4}
\end{center}
\end{figure}

\subsubsection{Horizontal segregation in `UW \& Gas' state}

Note in Fig.~\ref{fig:fig3}($a$-$c$) that most of the steel balls are in a gas-like state,
while the cluster and the UW are dominated by  glass balls, representing a state of {\it horizontal segregation}.
Since it is difficult to  quantify segregation in present experiments due to  the finite depth (more than five particle diameters) of our quasi-2D container,
we performed  event-driven simulations  of the same setup to understand the role of segregation on observed patterns.
The degree of segregation, along the horizontal and vertical directions, is quantified using the following order parameter~\citep{Rivas2011}:
\begin{equation}
   \delta = 1 - {\sum_i n_g(i) n_s(i)}/
     {\sqrt{\sum_i n^2_g(i) \sum_in^2_s(i)}},
     \label{eqn:eqn3}
\end{equation}
where $n_g$ and $n_s$ are the number density fields of glass and steel balls, respectively, in a given direction, and the sum runs from $0$ to
the corresponding system length in steps of one particle diameter.

The evolution of $\delta$, Eq.~(\ref{eqn:eqn3}), in the `UW \& Gas' state
is shown in Fig.~\ref{fig:fig4}($a$) for initial (main panel) and long (left inset) times, with parameter values as in  Fig.~\ref{fig:fig3}($b$). 
It is observed that the vertical segregation remains small at all times and the system evolves towards a state with significant horizontal segregation at late times.
The snapshots in Figs.~\ref{fig:fig4}($b$,$c$) at different times  confirm that the horizontal segregation precedes
the formation of an UW-state that coexists with a granular gas.
The  long-time simulation patterns are displayed in Figs.~\ref{fig:fig4}(c,d,e) at time-instances separated by $\tau/2$; the wavy-part  indeed represents an $f/2$ subharmonic-wave
that coexists with a disordered gas; notably, Figs.~\ref{fig:fig4}(c,e) bear striking resemblance to experimental pattern  in Fig.~\ref{fig:fig3}($b$). 
Overall,  the `UW+Gas'-patterns obtained from both experiment and simulation  consist of
a gas-like state coexisting with a period-2 subharmonic wave;
the simulations further confirmed that the primary  compositional characteristic of this  mixed pattern is  the horizontal-segregation of two species.

The right inset in Fig.~\ref{fig:fig4}($a$) clarifies an important point, the granular energy is unequally partitioned~\cite{FM2002} --
the heavier steel particles possess a higher kinetic energy ($E_s/E_g\approx 3$) and hence are in a more
mobile gas-like state; in contrast, the lighter glass particles possess a lower kinetic energy  and are thus relatively less mobile and they tend to move together, 
leading to a cluster of glass-rich particles.  The energy non-equipartition in a granular mixture could be a key factor for 
the observed segregation along the horizontal direction and the resulting phase-coexisting patterns.
A  driving mechanism  based on energy non-equipartition has been advocated for horizontal segregation observed
in simulations of a sub-monolayer binary mixture~\citep{Rivas2011} under vertical confinement,
although  in the latter system the resulting state appears to be akin to the gas-solid coexistence phenomenon.
On the other hand, at large shaking intensities $\Gamma\sim O(10)$ of present experiments, certain Knudsen-driven rarefied effects
are likely to be important, leading to anisotropic diffusion and consequently to horizontal segregation --
a model for such diffusive-mechanism is discussed in Sec.~IV.A.

\subsubsection{Possible stability mechanism of partial UW-state}

The ``{\it UW \& Gas}'' state, such as in Figs.~\ref{fig:fig3}($b,c$), is truly exotic since its long-time stability defies basic physics knowledge as we explain below.
For mono-disperse systems, the genesis of sub-harmonic UWs [e.g.~ in    Fig.~\ref{fig:fig3}(c)]
has been explained as a bifurcation from the BB-state due to the bending resistance of an effective elastic bar~\citep{Sano}.
More specifically,  when a compact layer of granular materials (the BB-state, constrained by two  side-walls)  impacts on the base plate,  the lowest layer of particles undergoes
dilation~\citep{Sano} which creates an effective  tension that increases with increasing $\Gamma$,
thus giving birth to a buckled state of the granular layer, such as   Fig.~\ref{fig:fig3}($c$), beyond some threshold $\Gamma$.
This  analogy with the buckling of an elastic-rod  is difficult to reconcile  with our finding of
`partial' UWs  (Fig.~\ref{fig:fig3}$b$) since the compact granular layer is free to dilate near the end where it is in constant touch with a granular gas;
moreover, the `unconstrained' dry granular materials cannot sustain tensile force.
For the same reason, another possible mechanism due to 
the sub-harmonic parametric instability~\cite{DFL1989,AT2006} can be ruled out for the genesis of  `partial' undulatory waves.

A closer analysis of movies (Movie-1) suggests a more complex process: when a partial UW hits the base, it dilates, resulting in  expulsion of particles to the gaseous region; 
however, there is  continual addition of `saltating' particles from  the gaseous region to UW-state -- 
both processes balance each other, leading to a stable `UW \& Gas' state.
We speculate that the collisional pressure within the  gaseous region may
also play the role of a dynamic barrier,  thereby helping the `partial' UWs to last for a long time.

\begin{figure}
\centering
(a)\includegraphics[width=3.2in]{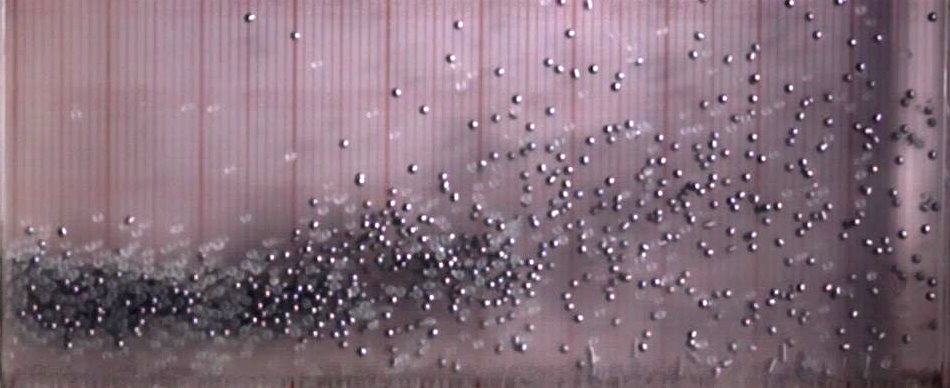}\\
(b)\includegraphics[width=3.2in]{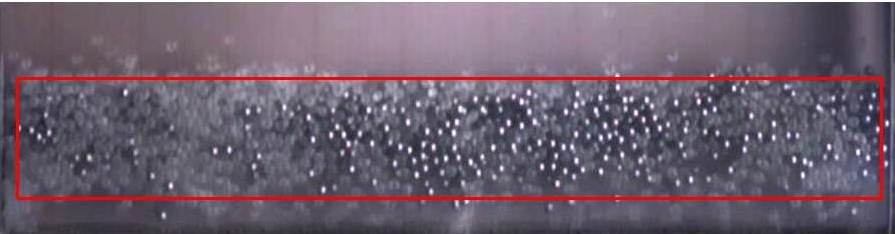}\\
(c)\includegraphics[width=3.2in]{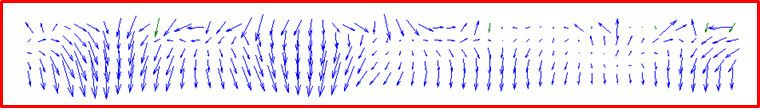} 
\caption{
Snapshots of granular Leidenfrost states (LS) and horizontal segregation:
(a) `{\it LS  \& Gas}' at $\Gamma= 50.22$ ($f=64.5$ Hz)  and $F=4$ and (b) complete LS at   $\Gamma=50.7$ ($f=65$ Hz) and $F =6$.
(c) Coarse-grained PIV velocity field for the boxed-region in panel $b$.
}
\label{fig:fig5}
\end{figure}

\begin{figure}
\centering
\includegraphics[width=3.6in]{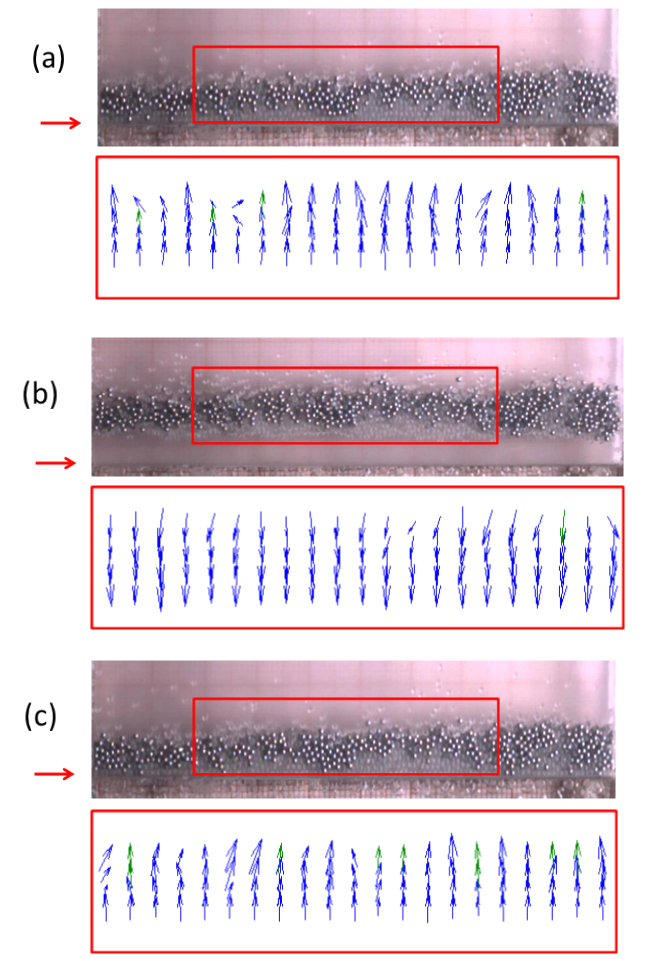}
\caption{
Snapshots and PIV-velocity fields of bouncing bed (BB) states at $\Gamma=4$ and $F=4$:
(a) $t=0$, (b) $\tau/2$ and (c) $\tau$, with $\tau=1/f$ being the time-period of harmonic shaking.
The horizontal arrow in each snapshot indicates the location the bottom of the box.
}
\label{fig:fig6}
\end{figure}

\subsection{Leidenfrost-like states and segregation} 

 We now turn our attention to higher shaking intensities ($\Gamma> 20$, not shown in Fig.~\ref{fig:fig1}), 
where the system goes through  different states with  increasing filling depth ($F$).   At $\Gamma=50$, there is a  gaseous state  for $F\leq 3$ and
this gives birth to a mixed state,  Fig.~\ref{fig:fig5}($a$) at $F=4$,  of a granular gas and a cluster on the right and the left of the container, respectively.
We verified that the cluster is in a Leidenfrost-like  state (LS)~\citep{Eshuis2005,Eshuis2007}
that corresponds to a density-inversion wherein a dense cloud of particles floats over a relatively dilute gaseous region of fast moving particles adjacent to the base.
(This analogy of a floating granular-layer with the original Leidenfrost-effect~\cite{Leidenfrost}
 of a liquid-drop  hovering over its own vapour-cushion was suggested by Eshuis {\it et al.}~\cite{Eshuis2005}.)
It has been established recently~\citep{AA2016} that the floating dense-layer has a liquid-like structure as confirmed via an analysis of  its pair-correlation function,
and the LS moves synchronously  with the vibrating container and hence this represents a period-1 pattern.
Therefore, the mixed state of ``Gas \& LS''  in  Fig.~\ref{fig:fig5}($a$)  represents the   coexistence of a disordered granular gas and a synchronous `liquid-like' period-1 wave.

A complete LS spanning the whole length of the box appears at larger filling depths (for $F>5.5$) as shown in  Fig.~\ref{fig:fig5}($b$).
As in the case of  ``Gas \& LS'' pattern, the steel and glass balls appear to be segregated along the horizontal direction in the LS pattern too --
this has been verified in simulations (see below).
The coarse-grained PIV velocity field (calculated by considering two successive frames separated by 1 ms)
within the boxed-region of Fig.~\ref{fig:fig5}($b$) is shown in Fig.~\ref{fig:fig5}($c$) -- there are strong
correlated motions along both horizontal and vertical directions. 
Such correlated motion constitutes one characteristic feature of the Leidenfrost state that distinguishes it from the BB-state~\cite{AA2016}
as it is evident from a comparison of Fig.~\ref{fig:fig5}($c$) with the velocity-vector plots in Fig.~\ref{fig:fig6}.
Figure~\ref{fig:fig6} shows three snapshots of the bouncing-bed state at times $t=0$, $\tau/2$ and $\tau$ -- 
each raw image of particle configuration is accompanied by its instantaneous PIV-velocity field that has been evaluated within the red-box.
It is seen that there is negligible lateral correlation of velocity
and the particle motion is primarily correlated with the vertical motion of the box as expected for a BB-state.

\begin{figure}
\centering
(a)\includegraphics[width=3.2in]{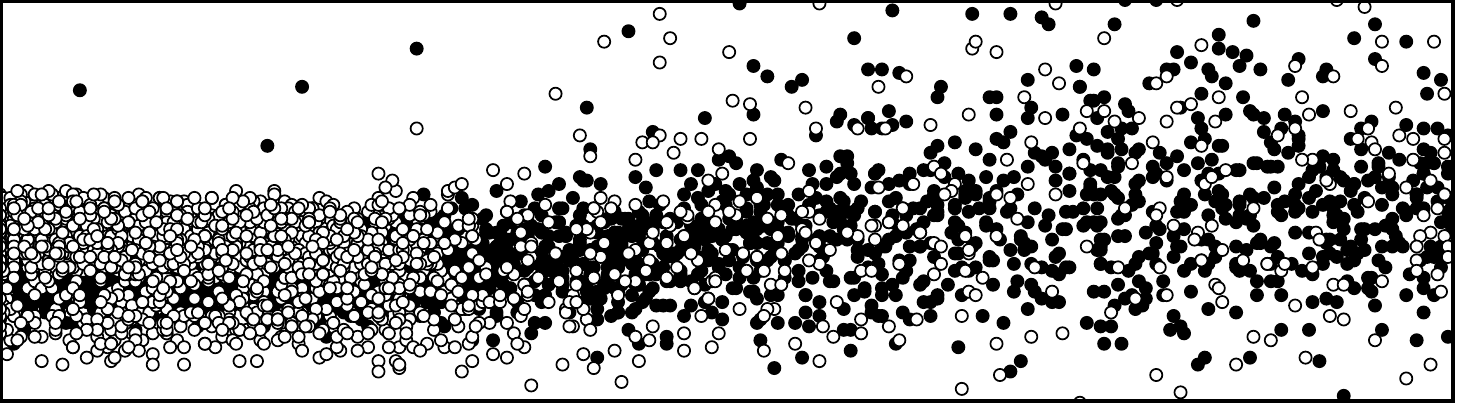}\\
(b)\includegraphics[width=3.2in]{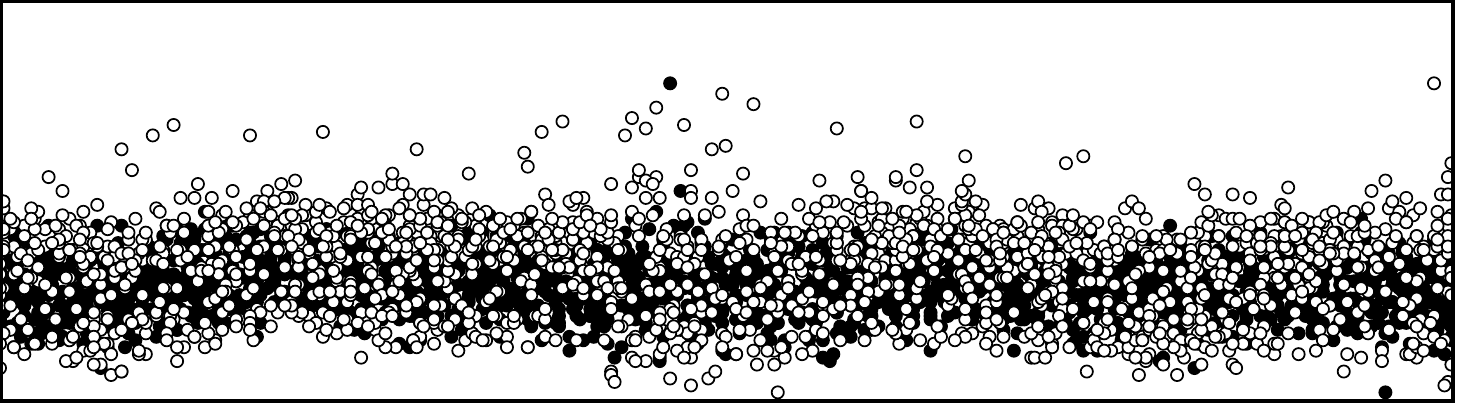}\\
(c)\includegraphics[width=3.4in]{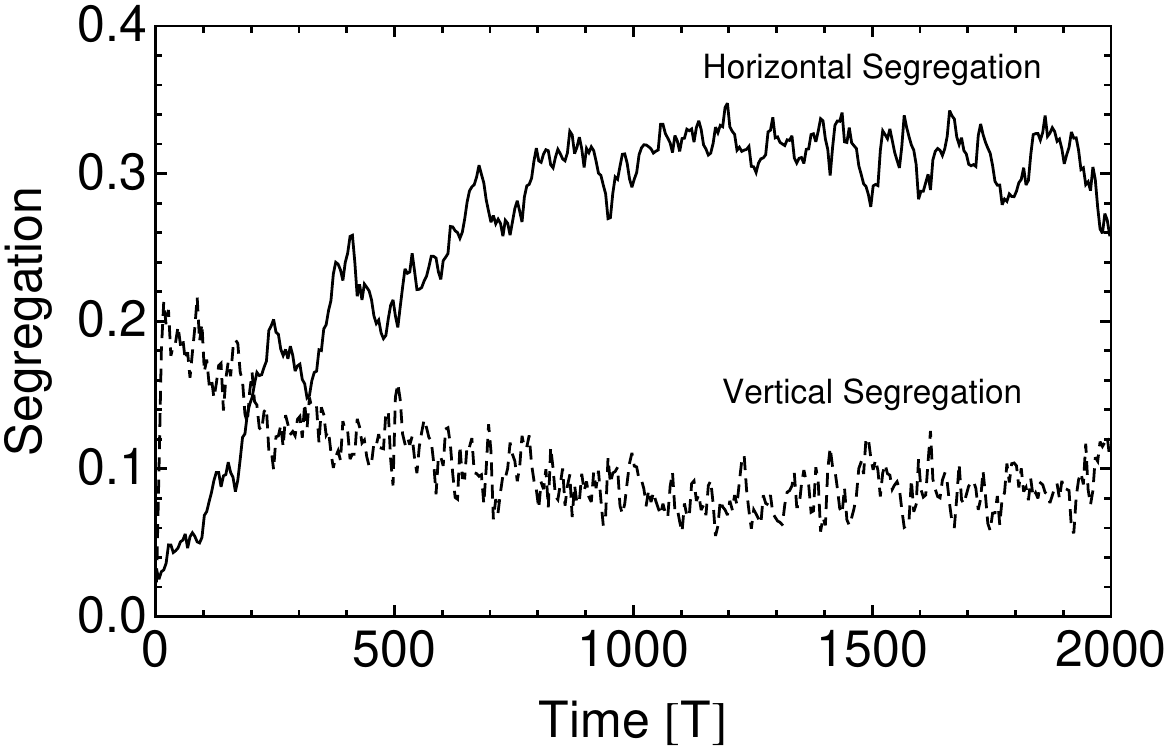} 
\caption{
Simulation snapshots of granular Leidenfrost states (LS) and horizontal segregation:
(a) `{\it LS  \& Gas}' at $F=4$ and (b) complete LS at   $F =6$;
other parameter values are as in Figs.~\ref{fig:fig5}(a) and \ref{fig:fig5}(b), respectively;
 filled and open circles refer to steel and glass balls, respectively.
(c) Simulation data for the evolution of horizontal (upper solid line) and vertical (lower dashed line) segregation indices for  parameter values of panel~$b$.
}
\label{fig:fig7}
\end{figure}

The simulation analog of Figs.~\ref{fig:fig5}(a,b) are displayed in Figs.~\ref{fig:fig7}(a,b), respectively.
As in experiments (Fig.~\ref{fig:fig5}a), the simulations confirmed that the partial Leidenfrost state (Fig.~\ref{fig:fig7}a)
is a mixed pattern of a gas-like state dominated by heavier steel balls that coexists with a glass-rich Leidenfrost-like state.
A closer inspection of  Fig.~\ref{fig:fig7}(b)  indicates that there is  segregation of steel and glass balls along the horizontal direction for the case of `complete' Leidenfrost state too.
For the latter case, the temporal evolution of segregation index, Eq.~(\ref{eqn:eqn3}), obtained from simulations is displayed in Fig.~\ref{fig:fig7}(c). 
There is clear  vertical segregation for a short time ($t/\tau < 200$), but 
the system eventually reaches a steady state with significant horizontal segregation. 
We therefore conclude that both the partial and complete Leidenfrost states are characterized
by the segregation of steel and glass balls along the horizontal direction.

\begin{figure}
\centering
(a) \includegraphics[width=3.0in]{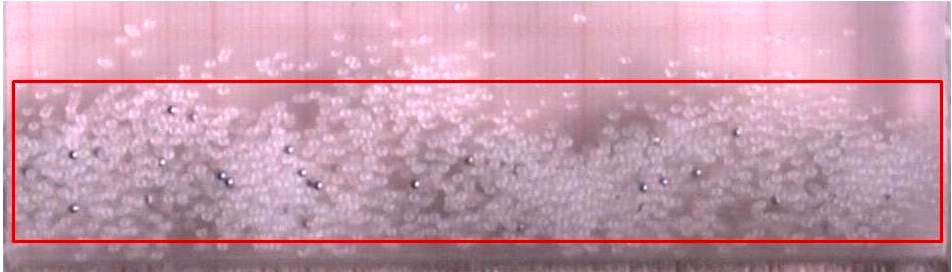}\\
\hspace*{0.6cm}\includegraphics[width=3.0in]{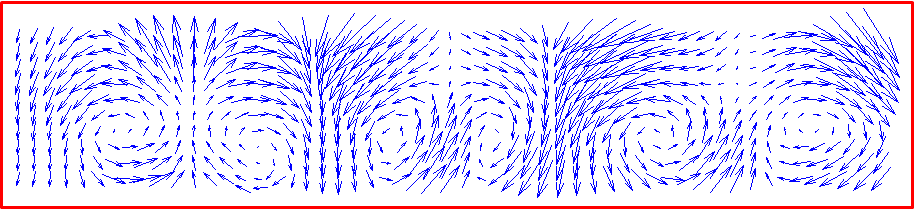}\\
(b)\includegraphics[width=3.0in]{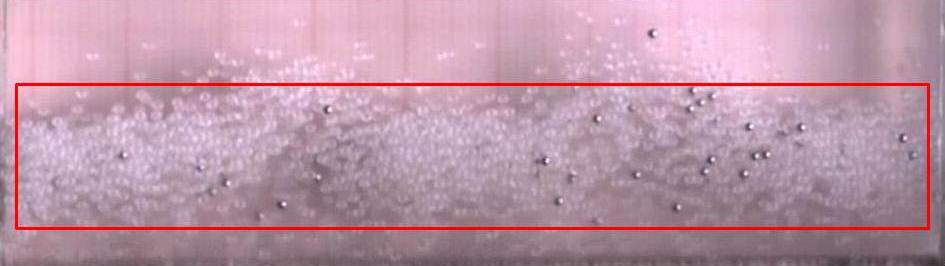}\\
\hspace*{0.5cm}\includegraphics[width=3.0in]{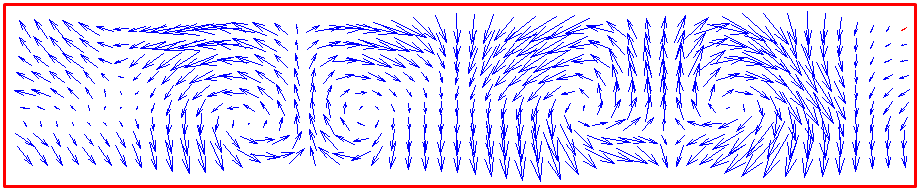}\\
(c)\includegraphics[width=3.0in]{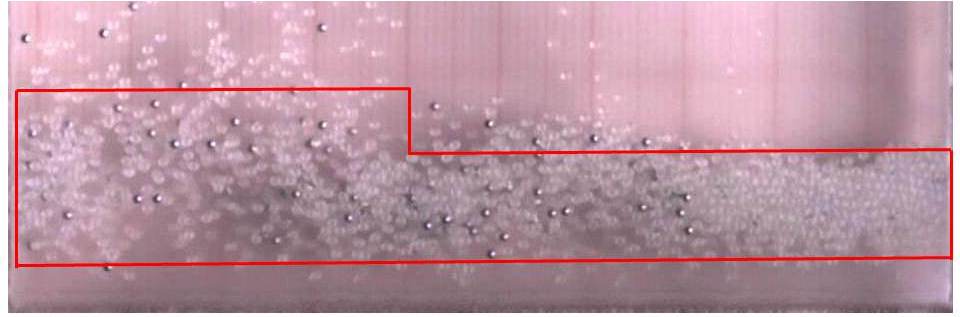}\\
\hspace*{0.5cm}\includegraphics[width=3.0in]{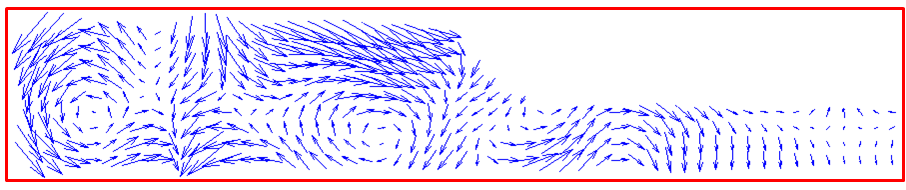}\\
(d)\includegraphics[width=3.0in]{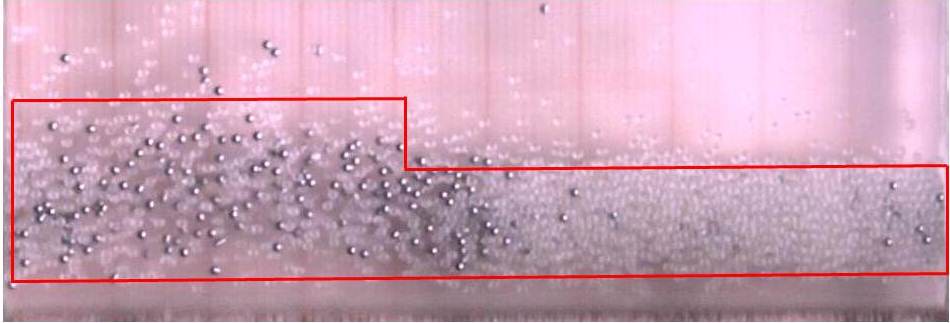}\\
\hspace*{0.6cm}\includegraphics[width=3.0in]{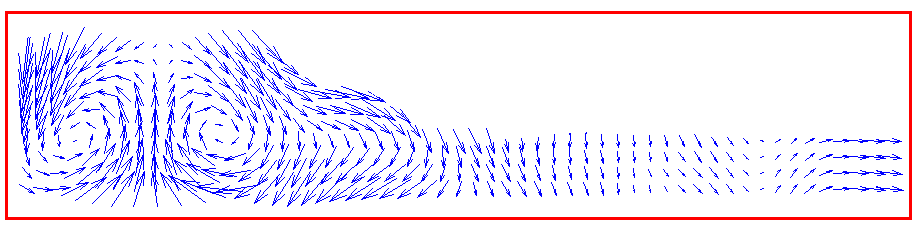}
\caption{
Instantaneous snapshot and its coarse-grained PIV velocity field  for mixtures of glass-steel balls with (a) $1\%$, (b) $2\%$,   (c) $5\%$  and (d) $10\%$ steel balls 
for  $\Gamma\approx 50$, $A/d=3$ and  $F=F_g+F_s =6$.  
The velocity field (lower panel) is calculated within the boxed-region (upper panel) of each snapshot.
}
\label{fig:fig8}
\end{figure}

\subsection{Granular convection and its control}  

For the same shaking intensity of $\Gamma\approx 50$  as in the Leidenfrost state in  Fig.~\ref{fig:fig5}(b),
the corresponding mono-disperse system of glass balls showed a convective motion~\cite{Eshuis2007},  containing six counter-rotating rolls 
which can be visualized from the supplementary Movie-2.
This indicates a strong influence of steel particles on the dynamics of the mixture. To better understand this, we progressively varied the concentration of steel particles.
With the addition of $1\%$ steel balls ($F_s/F=0.01$), the convection pattern and the number of rolls remain the same;
this is evident in  the PIV velocity field shown in the lower panel of Fig.~\ref{fig:fig8}($a$), with its  upper panel representing  the corresponding raw-image of particle configuration.
The velocity has been calculated within the red-boxed region of the raw-image --
this represents an instantaneous velocity field, calculated over two frames separated by 1 ms; 
however, due to the small number of particles in the system, the accuracy of the calculated velocity field is limited.
Here we are interested only in the gross features of the hydrodynamic velocity field, i.e., whether it contains a circulating motion or not.
It may be noted that the convection rolls, such as those in  Fig.~\ref{fig:fig8}($a$), represent the granular analog of the well-studied `buoyancy-induced'  thermal convection  
which was unequivocally  demonstrated first in experiments of Eshuis et al.~\cite{Eshuis2007}.

Figures~\ref{fig:fig8}(b-d) show a surprising effect --  the number of rolls decreases to four at $F_s/F=0.02$  (Fig.~\ref{fig:fig8}$b$) and 
to two at $F_s/F\geq 0.05$ (Figs.~\ref{fig:fig8}$c$ and \ref{fig:fig8}$d$).
In the latter two cases, a `complete' convection state, spanning the whole length of the container, ceases to exist
and the system degenerates into  a {\it partial} convection state,  characterized by a pair of counter-rotating rolls in one side of the  container 
and a Leidenfrost state on the other side.  
An inspection of raw-images in Fig.~\ref{fig:fig8}  indicates  that the convective rolls in binary mixtures are  populated
by steel balls, and the remaining part is dominated by lighter glass balls which represents a liquid-like (see the velocity field on
the right-side of each image in panels $c$ and $d$) Leidenfrost state.  The inspection of supplementary Movie-2 reveals that
 the convective motion within a partial-convection state is spatially non-uniform and unsteady, coupled with occasional `jet-like' expulsion of particles.
The present discovery of partial-convection states, such as Figs.~\ref{fig:fig8}(c,d), firmly establishes that
the ``side-walls''~\citep{GHS1992} are not required for the onset of  ``buoyancy-driven''  ~\citep{RRC2000,WHP2001,Paolotti2004,Eshuis2007,AA2016} granular convection.

\begin{figure}
\centering
\includegraphics[width=3.6in]{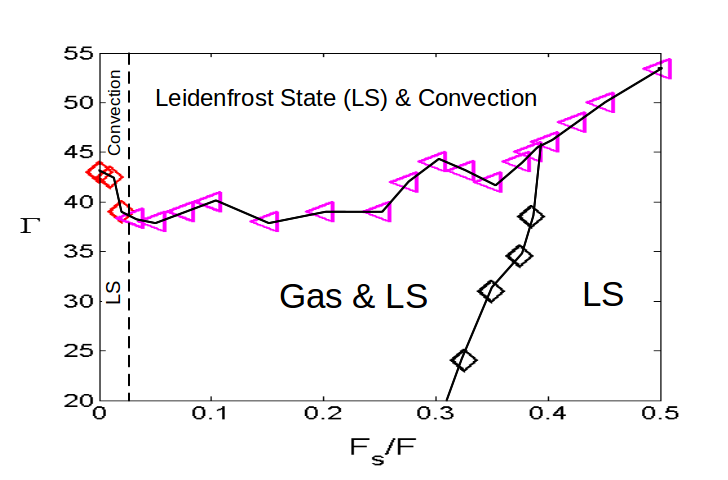}
\caption{
Effect of relative number fraction of steel balls  on the phase-diagram of patterns in ($\Gamma, F_s/F$)-plane for $F=F_g+F_s=6$ and $A/d=3$; 
other details are as in Fig.~\ref{fig:fig1}.  The vertical dashed line (at $F_s/F\approx 0.04$) represent an approximate phase-boundary between
two states (``Convection'' and ``LS \& Convection''; ``LS (Leidenfrost State)'' and ``Gas \& LS'').
}
\label{fig:fig9}
\end{figure}

The effect of relative number fraction ($F_s/F$) on convection is summarized  in Fig.~\ref{fig:fig9}.
For a specified shaking intensity  (say, $\Gamma=50$ in Fig.~\ref{fig:fig9}), the complete convection is possible only for a very small range of
$F_s/F\leq 0.04$, the partial-convection states occur for $F_s/F\in (0.05, 0.45)$ which eventually degenerate into  
a complete Leidenfrost state for $F_s/F>0.45$ (see Fig.~\ref{fig:fig5}$b$ for an equimolar mixture).
The red-diamonds symbols on the left of figure~\ref{fig:fig9} indicates that
the onset value of $\Gamma$ for convection decreases sharply with increasing steel balls ($F_s/F$) up-to $F_s/F \leq 0.04$ and increases thereafter:
therefore, the heavier particles play  a {\it dual} role of  enhancing and suppressing the convection intensity.
The delayed convection beyond a critical value of $F_s/F$ ($\geq 0.04$) is presumably due to the non-uniform fluidization of two-species.
The issue of non-uniform fluidization can be tied to the unequal partition of granular energy between steel and glass balls ($E_s\neq E_g$) 
as confirmed in present simulations too.
In summary (i) the multitude of pattern-transitions (among various mixed states) appearing from the rest-state 
of a solid-bed with increasing $\Gamma$ and varying total fill-height $F$, (ii) the partial convection state
and (iii) the recipe for convection-control by  varying the relative number-fraction of two species were not reported in previous experiments.

The possibility of convection-suppression  and its underlying mechanism have been  reported in a simulation study~\citep{Paolotti2004} with binary granular mixtures
to which we compare our results below.
In particular these authors~\citep{Paolotti2004} showed that adding about $12\%$ of quasi-elastic particles (of species-$2$ with its restitution coefficient $e_{22}=0.9992$) 
in an otherwise inelastic system  (of species-$1$ with $e_{11}=0.96$) can completely suppress the convection-rolls to a gas-like state 
for a specified shaking intensity (see their Figs.~11 and 12),  pointing towards the crucial role of  inelastic dissipation.
Note that their simulations correspond to extremely high shaking intensity $\Gamma=1250$ with  an  initial fill-height of $F\approx 19.5$ 
(and the mean volume fraction  of particles is about $0.0008$) and they did not report either  partial convection  or   Leidenfrost-like states or horizontal segregation.
In our experiments, the restitution coefficients of steel and glass balls are nearly equal ($e_{ss}\approx e_{gg}\approx 0.9$), and hence the present
transition-scenario is  unlikely to be driven solely by the difference in inelasticities of two species,
rather the combined effects of inelasticity and mass-disparity in an otherwise ``non-dilute'' binary mixture~\cite{FM2002,AL2002,AL2005}
is expected to lead to non-uniform fluidization of two species which might be  responsible for
the observed transition-route of  ``$Convection\to partialConvection\to LS$'' at fixed $\Gamma$ ($> 40$) with increasing $F_s/F$ (see Fig.~\ref{fig:fig9}).
Juxtaposing the  ``$Convection\to Gas$'' transition of Ref.~\citep{Paolotti2004} with Fig.~\ref{fig:fig9} at large enough $\Gamma$, 
we note the work of Ref.~\citep{Paolotti2004} belongs to a  ``dilute''  granular gas ($\Gamma=1250$)  for which we have no experimental data ($\Gamma < 55$) to compare with.
Nevertheless, the simulation transition can be explained by considering an effective restitution coefficient 
[i.e.~$e_{eff}=(e_{11}+e_{22})/2$ for an equimolar mixture] for the binary-mixture
such that the control parameter [i.e.~the inelasticity parameter $qN=N(1-e_{eff})/2$~\cite{RRC2000} which is equivalent to the Rayleigh number and hence
acts as a driving force for the onset of convection from a gas-like state] decreases with the addition of quasi-elastic particles.
Therefore, the convection-suppression mechanism of Ref.~\citep{Paolotti2004} is tied with the inelasticity-driven
 transition in a granular gas  (under some temperature gradient) where a gas-like state gives rise  to convection rolls with increasing inelasticity~\cite{RRC2000,KM2003} 
 and vice versa --  this constitutes a fundamental difference of  Ref.~\citep{Paolotti2004}  with our finding of  ``$Convection\to partialConvection\to LS$'' with increasing
 the number-density of a second-species at fixed $\Gamma$.

\section{Discussion and Conclusion}

\subsection{Spontaneous horizontal segregation and anisotropic diffusion}

The present experiments and simulations confirmed  that most  patterns  at large shaking intensities (Figs.~\ref{fig:fig2}-\ref{fig:fig5}, \ref{fig:fig7}-\ref{fig:fig8})
are characterized by the segregation of steel and glass balls along the horizontal direction,
with the heavier steel balls being in a gas-like disordered or a  convective state and the lighter glass balls in liquid-like UW/cluster/Leidenfrost state.
At this point we must emphasize that our finding of `spontaneous' horizontal segregation [e.g.~Fig.~\ref{fig:fig3}(a-c)] under vertical vibration 
is distinctly different from those of Ref.~\cite{SMZ2009}.
In the latter and related works,  the base of the container had an asymmetric `saw-tooth' shape 
which  acts like a ratchet for horizontal transport, resulting in `preferred' segregation along the horizontal direction.
In contrast, the present finding of horizontal segregation is not boundary-driven, rather it appears spontaneously
under  vertical excitations when the shaking intensity is large enough [$\Gamma\sim O(10)$ or more] and the fill-height is small enough [$F\sim O(5)$ or less].

\begin{figure}
\centering
(a)
\includegraphics[width=3.2in]{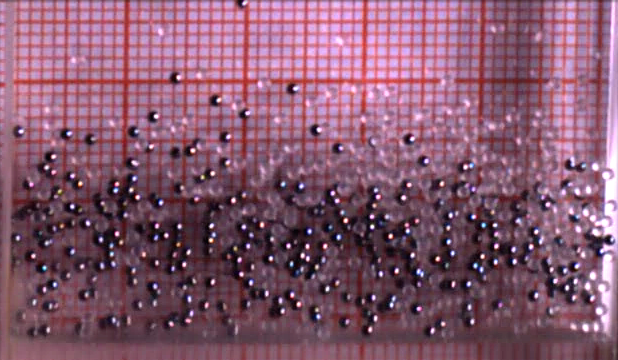}\\
(b)
\includegraphics[width=1.3in]{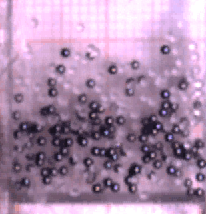}
\caption{
Effect of container length on patterns: (a) $L/d=50$ (a horizontally-segregated state, with a `glass-rich' cluster is seen on the right of the container) and (b) $L/d=20$
 (a gas-like nearly-homogeneous state) for $\Gamma=7$, $F_g=F_s=1.5$ and $A/d=3$.
}
\label{fig:fig10}
\end{figure}

If the segregation of two species along the horizontal direction is  responsible for observed coexisting patterns (such as Fig.~\ref{fig:fig3}),
the length ($L/d$) of the container must be sufficiently large to allow horizontal segregation to take place, otherwise
the two species are  expected to be well-mixed along the non-driven (horizontal) direction. This is indeed the case as demonstrated in Fig.~\ref{fig:fig10}.
While the larger container ($L/d=50$) in Fig.~\ref{fig:fig10}(a) represents a horizontally-segregated state,
the smaller container ($L/d=20$)  in Fig.~\ref{fig:fig10}(b) is a gas-like state with no visible horizontal segregation; for same parameter values of Fig.~\ref{fig:fig10}, 
increasing the container length to $L/d=100$ leads to another coexisting pattern of ``UW \& Gas'' state (see Fig.~\ref{fig:fig1}).
It is therefore clear that  the length of the Heleshaw-cell must be large enough for the appearance of phase-coexisting patterns (with two species being
segregated along the horizontal direction) as found in Fig.~\ref{fig:fig1}.

  Here we argue that certain Knudsen-driven rarefied effects   become  important
with increasing shaking intensity $\Gamma$, leading to anisotropic diffusion 
which can drive horizontal segregation under purely vertical shaking as we illustrate below;
our proposed mechanism for horizontal segregation is  based on well-known facts about  rarefied gas kinetic theory~\citep{Burnett1935,Grad1949,Kogan1969,CC1970}.
In a driven binary mixture, the net diffusion velocity between two species $A$ and $B$ is given by~\cite{GS1967,CC1970}
\begin{equation}
 {\boldsymbol v}^{AB} = {\boldsymbol v}^A - {\boldsymbol v}^B
    = - \frac{n}{n^An^B} {\boldsymbol D}^{AB}{\cdot}\left[ 
        \alpha{\boldsymbol\nabla}n + \beta{\boldsymbol\nabla}T
     \right] ,
     \label{eqn:vAB1}
\end{equation}
where $n=n^A + n^B$ is the number density of the mixture, and the expressions for $\alpha$ and $\beta$ can be obtained from kinetic theory;
the most general form of the  diffusion tensor (in two-dimensions) is 
\begin{equation}
   {\boldsymbol D}^{AB} =
   \left(
   \begin{array}{cc}
  {\mathcal D}_{xx} & {\mathcal D}_{xy} \\
  {\mathcal D}_{yx} & {\mathcal D}_{yy}
   \end{array}
   \right) .
   \label{eqn_D1}
\end{equation}
At Navier-Stokes-order (when the gradients of hydrodynamic fields are small, which is characterized by ``small'' values of the Knudsen number, $Kn\sim 0$), 
it is well-known that ${\mathcal D}_{xy}=0={\mathcal D}_{yx}$ and ${\mathcal D}_{xx}={\mathcal D}_{yy}={\mathcal D}_0$ 
and the scalar diffusion coefficient of a binary gas is given by~\citep{GS1967,CC1970}:
\begin{equation}
  {\mathcal D}_0 = \frac{1}{2 n d^{AB} } \left( \frac{m^{AB} T}{2\pi m^A m^B}\right)^{1/2} \propto \sqrt{T},
 \end{equation}
where $d^{AB}=d^A + d^B$, $m^{AB}=m^A + m^B$, and the $T$ is the mixture granular temperature.

When the Knudsen number deviates significantly from zero, the higher-order gradients become important which 
can be described  by Burnett-order theories~\citep{Burnett1935,Grad1949}.
Drawing an analogy between mass-transport and heat transport,  the form of the anisotropic diffusion tensor, Eq.~(\ref{eqn_D1}), can be justified~\citep{SA2014},
 with $({\mathcal D}_{xy}, {\mathcal D}_{yx})\neq 0$ and ${\mathcal D}_{xx}\neq {\mathcal D}_{yy}$; it has been established~\cite{SA2014} that the off-diagonal terms of the 
 conductivity/diffusion  tensor appear at the second-order in a gradient expansion [i.e. at $O(Kn^2)$].
 Assuming that the anisotropic diffusion tensor is given by Eq.~(\ref{eqn_D1}) and 
under purely vertical shaking along $y$-direction as in our experiments, the horizontal component of the diffusion velocity follows from Eq.~(\ref{eqn:vAB1}):
\begin{equation}
   v^{AB}_x = \alpha {\mathcal D}_{xy} \frac{\partial n}{\partial y} + \beta {\mathcal D}_{xy} \frac{\partial T}{\partial y}.
 \end{equation}
It is clear that the $v_x^{AB}$ is non-zero if and only if  the off-diagonal components of the diffusion tensor are non-zero 
and therefore the vertical-gradients alone (in density and/or temperature) can drive segregation of two species along the horizontal direction.
This provides a tentative explanation for the observed `spontaneous' horizontal segregation in our setup -- a  detailed theory can be worked out once
all components of the Burnett-order diffusion tensor and other related transport coefficients are known which is left to a future work.

\subsection{Conclusions and outlook}

Using both experiments and simulations, we uncovered new patterns in vertically vibrated binary granular mixtures that are
characterized by the coexistence of two phases (Movie 1):
(i) a disordered/asynchronous granular gas coexisting with (ia) a subharmonic (period-2) undulatory wave,
(ib) a synchronous (period-1) cluster, and (ic) a  Leidenfrost-like  state (of period-1~\cite{AA2016});  
and (ii) a Leidenfrost state coexisting with a pair of rolls (i.e. ~a `partial' convection state).
The coexistence of time-periodic synchronous liquid-like and disordered/asynchronous gas-like states is dubbed `phase-coexisting' patterns --
these mixed patterns  are made of two parts having different temporal and spatial symmetries.
Both experiments and simulations confirmed that the onset of all mixed patterns
is a consequence of the horizontal segregation of heavier and lighter particles.  
We discovered a giant effect of adding a small amount of heavier particles that can control (i) the onset of  granular convection,
(ii) the number of convection-rolls and (iii) the partial convection state (Movie 2).
On the whole, controlling patterns in granular flows using such a simple recipe may have far reaching consequences from the viewpoint of potential applications in processing industries.

The segregation between heavier and lighter particles along the horizontal direction is shown  to be  the key factor for all observed patterns as well as for the protocol for convection control.
A possible mechanism for `spontaneous' horizontal segregation under vertical vibration has been put forward in terms of an `anisotropic' diffusion tensor --
such a diffusive mechanism is likely to be operative at high-shaking intensities such that the heavier species stays in a relatively rarefied/dilute state
compared to the lighter-species.
Increasing the lateral confinement (i.e.~by decreasing the length, $L$, of the container)  leads to patterns in which two species appear well-mixed,
which suggests that the diffusion of two species  along the horizontal direction indeed plays a key role on the genesis of horizontal-segregation,
resulting in a variety of phase-coexisting patterns that we have uncovered.
The derivation of the exact form of the diffusion tensor and the underlying analyses are  beyond the scope of the present work and left to  future.

\vspace*{1.0cm}
\noindent{\bf Acknowledgements}\\
We are grateful for the funding support from the Department of Atomic Energy, Government of India  (DAE/MA/4265).
Experiments were carried out by I. H. Ansari and the numerical simulations by N. Rivas.
We sincerely thank Sunil  Bharadwaj for help with PIV analysis of some data, and Stefan Luding for  comments on the draft-manuscript.


\begin{thebibliography}{99}
\bibitem[Brown (1939)]{Brown1939}
  {L.~R. Brown},
  Fundamental principles of segregation,
  \emph{J.~Inst.~Fuel} {\bf 13}, 15-19 (1939).
\bibitem[Kudrolli (2004)]{Kudrolli2004}
  {A. Kudrolli},
  Size separation in vibrated granular matter,
   \emph{Rep.~Prog.~Phys.} \textbf{67}, 209-247 (2000).
\bibitem[Ottino \& Khakhar (2000)]{OK2000}
  {J.~M. Ottino and D.~V.  Khakhar},
  Mixing and segregation of granular materials,
  \emph{Ann.~Rev.~Fluid Mech.} \textbf{32}, 55-91 (2000).
\bibitem[Rosato et al. (1987)]{Rosato1987}
  {A. Rosato, K.~J.  Strandburg, F.  Prinz and R.~H. Swendsen},
  Monte Carlo simulation of particulate matter segregation,
  \emph{Phys.~Rev.~Lett.} {\bf 58}, 1038-1040 (1987).
\bibitem[Shinbrot \& Muzzio (1998)]{SM1998}
  {T. Shinbrot and F.~J. Muzzio}, 
   Reverse buoyancy in shaken granular beds,
  \emph{Phys.~Rev.~Lett.} {\bf 81}, 4365-4368 (1998).
\bibitem[Breu et al. (2003)]{Breu2003}
  {A.~P Breu, H.~M. Ensner, C.~A.  Kruelle, and I. Rehberg},
  Reversing the Brazil nut effect: Competition between percolation and condensation,
   \emph{Phys.~Rev.~Lett.} {\bf 90}, 014302 (2003).
\bibitem[Hong, Quinn \& Luding (2001)]{HQL2001}
  {D.~C. Hong, P.~V. Quinn, and S. Luding},
  Reverse Brazil nut problem: Competition between percolation and condensation,
   \emph{Phys.~Rev.~Lett.} {\bf 86}, 3423-3226 (2001).
\bibitem[Cl\'ement,  Duran \& Rajchenbach (1992)]{CDR1992}
  {E. Cl\'ement, J. Duran, and J. Rajchenbach},
  Experimental study of heaping in a two-dimensional sand pile,
  \emph{Phys.~Rev.~Lett.} {\bf 69}, 1189-1192 (1992).
\bibitem[Douady, Fauve \& Laroche (1989)]{DFL1989}
  {S. Douady, S. Fauve, and C. Laroche},
  Subharmonic instabilities and defects in a granular layer under vertical vibrations,
  \emph{Europhys.~Lett.} {\bf 8}, 621-626 (1989).
\bibitem[Faraday (1831)]{Faraday}
  {M. Faraday},
  On a peculiar class of acoustical figures; and on certain forms assumed by groups of particles upon vibrating elastic surfaces,
  \emph{Phil.~Trans.~R.~Soc.~Lond.} {\bf 52}, 299-318 (1831).
\bibitem[Umbanhower, Melo \& Swinney (1996)]{UMS1996}
  {P.~B. Umbanhower, F. Melo, and  H.~L. Swinney},
   Localized excitations in a vertically vibrated granular layer,
  \emph{Nature} \textbf{382}, 793-796 (1996).
\bibitem[Gallas,  Herrmann \& Sokolowski (1992)]{GHS1992}
  {J.~A.~C. Gallas, H.~J. Herrmann, and S. Sokolowski},
  Convection cells in vibrating granular media,
  \emph{Phys.~Rev.~Lett.} {\bf 69}, 1371-1374 (1992).
\bibitem[Wildman, Huntley \& Parker (2001)]{WHP2001}
  {R.~D. Wildman, J.~M.  Huntley,  and  D.~J. Parker},
  Convection in highly fluidized three-dimensional granular beds,
  \emph{Phys.~Rev.~Lett.} {\bf 86}, 3304-3307 (2001).
\bibitem[Eshuis \etal (2007)]{Eshuis2007}
  {P. Eshuis, K.  van der Weele, D. van der Meer, R. Bos, and  D. Lohse},
  Phase diagram of vertically shaken granular matter,
  \emph{Phys.~Fluids} {\bf 19}, 123301 (2007).
  \bibitem{Eshuis2010}
  P. Eshuis, D. van der Meer, M. Alam, H.~J. van Gerner, K. van der Weele and  D. Lohse,
  Onset of Convection in Strongly Shaken Granular Matter,
  \emph{Phys.~Rev.~Lett.} \textbf{104}, 038001 (2010).
\bibitem[Eshuis \etal (2005)]{Eshuis2005}
  {P. Eshuis, K.  van der Weele, D. van der Meer, and  D. Lohse},
  Granular Leidenfrost effect: Experiment and theory of floating particle  clusters,
  \emph{Phys.~Rev.~Lett.} {\bf 95}, 258001 (2005).
\bibitem[Meerson, P\"oschel \& Bromberg (2003)]{MPB2003}
  {B. Meerson,  T. P\"oschel, and Y. Bromberg},
   Close-packed floating clusters: granular hydrodynamics beyond a freezing point?
  \emph{Phys.~Rev.~Lett.} {\bf 91}, 024301 (2003).
\bibitem[P\"oschel \& Herrmann (1995)]{PH1995}
  {T. P\"oschel and  H.~J. Herrmann},
  Size segregation and convection,
  \emph{Europhys.~Lett.} {\bf 29}, 124-128 (1995).
\bibitem[Burtally, King \& Swift (2002)]{BKS2002}
   {N. Burtally, P.~J. King, and M.~R. Swift},
   Spontaneous air-driven separation in vertically vibrated fine  granular mixtures,
   \emph{Science} {\bf 295}, 1877 (2002).
\bibitem[Olafsen \&  Urbach (1998)]{OU1998}
  {J.~S. Olafsen and  J.~S. Urbach},
  Clustering, order, and collapse in a driven granular monolayer,
  \emph{Phys.~Rev.~Lett.} {\bf 81}, 4369-4372 (1998).
\bibitem[Aranson \& Tsimring (2006)]{AT2006}
  {I.~S. Aranson  and  L.~S. Tsimring},
  Patterns and collective behaviour in granular media: theoretical concepts,
   \emph{Rev.~Mod.~Phys} {\bf 78}, 641 (2006).
 \bibitem[Shukla et al. (2014)]{SAMLA2014}
  {P. Shukla, I.~H.  Ansari, D.  van der  Meer, D. Lohse,  and M. Alam}, 
  Nonlinear instability and convection in vertically vibrated granular bed,
  \emph{J.~Fluid Mech.} {\bf 761}, 123-167 (2014).
\bibitem[Ansari \& Alam (2016)]{AA2016}
  {I.~H. Ansari  and  M. Alam},
   Pattern transition, microstructure and dynamics in vertically vibrated granular bed,
   \emph{Phys.~Rev.~E}  \textbf{93}, 052901 (2016).
\bibitem[Ansari \& Alam (2013)]{AA2013a}
   {I.~H. Ansari and M. Alam},
   Exotic patterns and convection control in a vibrated bed of binary  granular mixtures,
    \emph{Bulletin of American Physical Society} (67th Annual Meeting of the APS Division of Fluid Dynamics), 
  vol.~\textbf{58}, Number 18 (Abstract Number: G.24.00008), (2013).
\bibitem[Ansari \& Alam (2013)]{AA2013}
  {I.~H. Ansari and  M. Alam},
  Patterns and velocity field in vertically vibrated granular materials,
  \emph{AIP Conf. Proceedings} (Editors: A.~Yu \etal.) \textbf{1542}, 775-778 (2013).
\bibitem[Rivas et al. (2011)]{Rivas2011}
 {N. Rivas, P. Cordero, D.  Risso, and  R. Soto},
  Segregation in quasi-two-dimensional granular systems,
 \emph{New J.~Phys.}  {\bf 13}, 055018 (2011).
\bibitem[G\"otzendorfer et al. (2006)]{GKRS2006}
  {A. G\"otzendorfer, C.~A. Kruelle, I. Rheberg,  and D. Svensek}, 
  Localized subharmonic waves in a circularly vibrated granular bed,
 \emph{Phys.~Rev.~Lett.} {\bf 97}, 198001 (2006).
\bibitem[Feitosa \& Menon  (2002)]{FM2002}
  {K. Feitosa and N. Menon},
  Breakdown of energy equipartition in a 2D binary vibrated granular gas,
   \emph{Phys.~Rev.~Lett.} {\bf 88}, 198301 (2002).
\bibitem[Sano (2005)]{Sano}
  {O. Sano},
  Dilatancy, buckling, and undulations on a vertically vibrating granular layer,
  \emph{Phys.~Rev.~E} \textbf{72}, 051302 (2005).
\bibitem[Leidenfrost (1756)]{Leidenfrost}
 {J.~G. Leidenfrost},
 \emph{De Aquae Communis Nonnullis Qualitatibus Tractatus} (University of Duisburg, Duisburg, Germany, 1756).
\bibitem[Ramirez, Risso, \& Cordero (2000)]{RRC2000}
  {R. Ramirez, D. Risso, and  P. Cordero},
  Thermal convection in fluidized granular systems,
  \emph{Phys.~Rev.~Lett.} {\bf 85}, 1230-1233 (2000).
\bibitem[Paolotti \etal (2004)]{Paolotti2004}
  {D. Paolotti, A. Barrat, U.~M.~B. Marconi, and A. Puglisi},
  Thermal convection in monodisperse and bidisperse granular gases: a simulation study,
  \emph{Phys.~Rev.~E} {\bf 69}, 061304 (2004).
 \bibitem[Alam \& Luding (2002)]{AL2002}
 {M. Alam and  S. Luding},
 How good is the equipartition assumption for the transport properties of a granular mixture?
 \emph{Granul.~Matter} \textbf{4}, 139 (2002).
\bibitem[Alam \& Luding (2005)]{AL2005}
 {M. Alam and  S. Luding},
  Energy nonequipartition, rheology and microstructure in sheared bidisperse granular mixtures, 
 \emph{Phys.~Fluids} \textbf{17}, 063303 (2005).
 \bibitem{KM2003}
  E. Khain and B. Meerson,
  Onset of thermal convection in a horizontal layer of granular gas,
  \emph{Phys.~Rev.~E}, \textbf{67}, 021306 (2003).
\bibitem[Shi, Miao \& Zhang (2009)]{SMZ2009}
  {X. Shi, G. Miao and H.  Zhang},
  Horizontal segregation in a vertically vibrated binary granular system,
  \emph{Phys.~Rev.~E} {\bf 80}, 061306 (2009)
 \bibitem{Burnett1935}
  D. Burnett,
  The distribution velocities in a slightly non-uniform gas,
  \emph{Proc.~Lond.~Math.~Soc.} {\bf 39}, 385-430 (1970).
\bibitem{Grad1949}
 H. Grad,
 On the kinetic theory of rarefied gases,
 \emph{Commun.~Pure  Appl.~Maths.} {\bf 2}, 331--407 (1949).
\bibitem{Kogan1969}
  M.~N. Kogan,
  \textit{Rarefied Gas Dynamics} (Plenum Press, New York, 1969).
 \bibitem{CC1970}
  S. Chapman and T.~G. Cowling,
  \textit{The Mathematical Theory for Non-Uniform Gases} (Cambridge University Press, 1970).
\bibitem{GS1967}
E. Goldman and L. Sirovich,
Equations for gas mixtures,
\emph{Phys.~Fluids} \textbf{10}, 1928-1940 (1967).
\bibitem{SA2014}
S. Saha and M. Alam, 
  Non-Newtonian stress, collisional dissipation and heat flux in the shear flow of inelastic disks: a reduction via Grad's moment method,
  \emph{J. Fluid Mech.} {\bf 757}, 251-296 (2014).
\end{thebibliography}
\end{document}